\documentstyle[preprint,aps]{revtex}
\begin{document}
\preprint{GTP-97-01}
\draft
\title{ Propagator for an Aharonov-Bohm-Coulomb system }
\author
{$^{a}$D.K.Park, $^{b}$Sahng-Kyoon Yoo, $^{c}$Soo-Young Lee, $^{c}$Jae-Rok Kahng, \\
$^{d}$Chang Soo Park, $^{e}$Eui-Soon Yim, and $^{f}$C.H.Lee }
\address
{$^{a}$ Department of Physics, Kyungnam University, Masan, 631-701, Korea \\ 
$^{b}$ Department of Physics, Seonam University, Namwon, 590-170, Korea \\
$^{c}$ Department of Physics, Korea University, Seoul, 136-701, Korea \\
$^{d}$ Department of Physics, Dankook University, Cheonan, 330-180, Korea \\
$^{e}$ Department of Physics, Semyung University, Chechon, 390-230, Korea \\
$^{f}$ D\&S Dept.,R\&D Center, Anam Industrial CO., LTD, Seoul 133-120, Korea}
\maketitle
\begin{abstract}
The propagator of three-dimensional Aharonov-Bohm-Coulomb system is
calculated by following the Duru-Kleinert method. It is shown that
the system is reduced to two independent two dimensional 
Aharonov-Bohm plus harmonic oscillator systems through dimensional
extension and Kustaanheimo-Stiefel transformation. The energy
spectrum is deduced.
\end{abstract}

\newpage
Since Aharonov and Bohm(AB) gave a physical significance to the vector
potential in 1959[1], there are lots of theoretical and 
experimental attempts[2] to establish the AB effect.
Recent attention given to the phenomenon is mostly related to the anyonic
theory[3] which tries to understand the high-$T_c$ 
superconductivity phenomena using fractional statistics. 
Since AB interaction is interaction between charge and magnetic flux, 
anyon-anyon interaction naturally requires Coulomb modification. 
In this letter
we will study on the AB problem with Coulomb modification by invoking the 
path-integral formalism in three dimensions. 

For a last decade, the two-[4] and
three-dimensional[5,6]
Aharonov-Bohm-Coulomb(ABC) systems have been studied by using the
various different methods. Especially Ref.[5] used the
path-integral formalism in the spherical coordinate and obtained the
energy-dependent Green's function and the bound-state energy spectrum
in ABC system. Same bound-state energy spectrum has been found in 
Ref.[6], in which the generalized ABC system is studied with an
aid of Kustaanheimo-Stiefel(KS) transformation[7]
by solving Schr\"{o}dinger equation.
The KS transformation is also used for the evaluation of the exact
propagator in hydrogen atom[8,9].

In this letter we will derive the exact propagator for ABC system by invoking
the Duru-Kleinert method. Let us start with 
Hamiltonian
\begin{equation}
\hat{H} = \frac{(\vec{p} - e \vec{A})^2}{2 M} + \frac{\xi}{r}
\end{equation}
where AB potential is 
\begin{equation}
\vec{A} = \frac{\alpha}{e} \frac{y \hat{x} - x \hat{y}}{r^2 - z^2}
\end{equation}
in Coulomb gauge.
Following Ref.[10], we can describe the fixed-energy amplitude $(\vec{x}_b
\mid \vec{x}_a)_E$ as a pseudotime-sliced path integral
\begin{eqnarray}
& &(\vec{x}_b \mid \vec{x}_a)_E  \\ \nonumber 
&=& \int_0^{\infty} ds f_r(\vec{x}_b) f_l(\vec{x}_a) \int D\vec{x}
  \int D\vec{p} exp \left[ i \int_0^s ds \
   \left[ 
         \vec{p} \cdot \vec{x}' - f_l(\vec{x}) (\hat{H} - E) f_r(\vec{x})
                                                                   \right]
                                                                  \right] 
\end{eqnarray}
where $f_l(\vec{x})$ and $f_r(\vec{x})$ are regulating functions defined
in chapter 12 of Ref.[10] and $\vec{x}' = d\vec{x} / ds$.
Pseudotime $s$ is defined by $ds / dt = f_l(\vec{x}) f_r(\vec{x})$.
It is worthwhile to note that fixed-energy amplitude has a relation with 
energy-dependent Green's function , which is Laplace transform of Euclidean
Kernel, as follows:
\begin{equation}
(\vec{x}_b \mid \vec{x}_a)_E = -i \hat{G}[\vec{x}_b, \vec{x}_a : -E].
\end{equation}
After choosing 
\begin{eqnarray}
f_l(\vec{x})&=& r^{1 - \lambda}  \\ \nonumber 
f_r(\vec{x})&=& r^{\lambda}
\end{eqnarray}
and performing $\vec{p}$-integration, one can derive the following form 
straightforwardly:
\begin{eqnarray}
& &(\vec{x}_b \mid \vec{x}_a)_E  \\ \nonumber 
&=& \lim_{N \rightarrow \infty} (N + 1) \int_0^{\infty} d \epsilon_s
   r_b^{\lambda} r_a^{1 - \lambda}                         
  \left( 
        \frac{M}{2 \pi i \epsilon_s r_a^{\lambda} r_b^{1 - \lambda}}
                                                 \right)^{3 / 2}
\int
             \left[ \prod_{j=2}^{N+1} 
                   \left( 
                         \frac{M}{2 \pi i \epsilon_s r_{j-1}} 
                                                             \right)^{3/2}
                   d \Delta \vec{x}_j
                                                           \right]
                   e^{i A_{0, E}^N}
\end{eqnarray}
where $(N+1) \epsilon_s = s_b - s_a \equiv s$ and
\begin{equation}
A_{0, E}^{N} = - \xi (N+1) \epsilon_s + 
               \sum_{j=1}^{N+1} 
                    \left[
                          \frac{M}{2} 
\frac{(\vec{x}_j - \vec{x}_{j-1})^2}
     {\epsilon_s r_j^{1 - \lambda} r_{j-1}^{\lambda}}
            + e \vec{A}_j \cdot (\vec{x}_j - \vec{x}_{j - 1} )
            + \epsilon_s E r_j
                               \right]
\end{equation}
whose contiuum limit is
\begin{equation}
A_{0, E}[\vec{x}] = - \xi s + \int_0^s ds 
                              \left[ \frac{M}{2 r} \vec{x}'^2 + e \vec{A} \cdot
                                     \vec{x}' + E r
                                                   \right].
\end{equation}
At this stage we have to exploit the dimensional extension
technique[11] for the incorporation of the KS transformation into the 
path integral formalism.
This is achieved by inserting the following trivial identity 
\begin{equation}
\int_{- \infty}^{\infty}
         \left[
               \prod_{j=1}^{N+1} \left(
                                       \frac{M}{2 \pi i \epsilon_s 
                                                r_j^{1 - \lambda} 
                                                r_{j-1}^{\lambda}}
                                             \right)^{1/2} d \Delta \eta_j
                                                        \right]
         exp \left[
                   i \sum_{j=1}^{N+1} \frac{M}{2}
                   \frac{\Delta \eta_j^2}{\epsilon_s r_j^{1 - \lambda}
                                                     r_{j-1}^{\lambda}}
                                  \right] = 1
\end{equation}
into Eq.(6).
Hence, $(\vec{x}_b \mid \vec{x}_a)_E$ is expressed as four dimensional
path integral
\begin{eqnarray}
(\vec{x}_b \mid \vec{x}_a)_E
&=& (N+1) \int_{0}^{\infty} d \epsilon_s r_b^{\lambda} r_a^{1 - \lambda}
          \left( \frac{M}{2 \pi i \epsilon_s r_a^{\lambda} r_b^{1 - \lambda}}
                                      \right)^2  \\ \nonumber 
           &\times& \int d \Delta \eta_1
                   \int \left[
                              \prod_{j=2}^{N+1} 
                              \left( \frac{M}{2 \pi i \epsilon_s r_{j-1}}
                                          \right)^2
                              d \Delta \vec{x}_j d \Delta \eta_j
                                           \right]
                  e^{i A_E^N}
\end{eqnarray}
where
\begin{equation}
A_E^N = - \xi (N+1) \epsilon_s + 
        \sum_{j=1}^{N+1} \left[ 
                               \frac{M}{2} \frac{\Delta \vec{x}_j^2 + \Delta 
                                                 \eta_j^2}{\epsilon_s 
                                                           r_j^{1 - \lambda}
                                                           r_{j-1}^{\lambda} }
                               + e \vec{A}_j \cdot \Delta \vec{x}_j
                               + \epsilon_s E r_j
                                                   \right].
\end{equation}
By using the the following approximation
\begin{eqnarray}
& &r_b^{\lambda} r_a^{1 - \lambda}
 \left( \frac{M}{2 \pi i \epsilon_s r_a^{\lambda} r_b^{1 - \lambda} }
                                      \right)^2
\prod_{j=2}^{N+1} \left(\frac{M}{2 \pi i \epsilon_s r_{j-1}} \right)^2 \\ 
\nonumber
&\approx& \left( \frac{M}{2 \pi i \epsilon_s} \right)^2 \frac{1}{r_a}
        \left[
              \prod_{j=2}^{N+1} \left( \frac{M}{2 \pi i \epsilon_s r_j} 
                                                    \right)^2 \right]
        exp \left[ 3 \lambda \sum_{j=1}^{N+1} \ln \frac{r_j}{r_{j-1}} \right]
\end{eqnarray}
one can change Eq.(10) into
\begin{eqnarray}
(\vec{x}_b \mid \vec{x}_a )_E &=& (N+1) \int_0^{\infty} d \epsilon_s
       \left( \frac{M}{2 \pi i \epsilon_s} \right)^2 \\ \nonumber
   &\times&    \int \frac{d \Delta \eta_1}{r_a}
       \int \left[ 
                  \prod_{j=2}^{N+1} \left( \frac{M}{2 \pi i \epsilon_s r_j}
                                                            \right)^2                                             
                   d \Delta \vec{x}_j d \Delta \eta_j
                                                      \right]
                   e^{i (A_E^N + A_f^N)}
\end{eqnarray}
where
\begin{equation}
A_f^N = - 3 i \lambda \sum_{j=1}^{N+1} \ln \frac{r_j}{r_{j-1}}.
\end{equation}
Since final continuum limit of $(\vec{x}_b \mid \vec{x}_a )_E$ is 
independent of $\lambda$ [10], we set $\lambda = 0$, which gives
\begin{equation}
(\vec{x}_b \mid \vec{x}_a )_E = (N+1) \int_0^{\infty} d \epsilon_s
\left( \frac{M}{2 \pi i \epsilon_s} \right)^2
\int \frac{d \Delta \eta_1}{r_a}
\int \left[ \prod_{j=2}^{N+1}
            \left( \frac{M}{2 \pi i \epsilon_s r_j}  \right)^2
            d \Delta \vec{x}_j d \Delta \eta_j                \right]
        e^{i A_{E, \lambda = 0}^N}.
\end{equation}

We now apply the KS transformation defined by
\begin{equation}
\left( \begin{array}{c}
  \Delta x \\ \Delta y \\ \Delta z \\ \Delta \eta
  \end{array} \right)
  = 2 A(\vec{u})
  \left( \begin{array}{c}
  \Delta u^1 \\ \Delta u^2 \\ \Delta u^3 \\ \Delta u^4
  \end{array} \right),
\end{equation}
where
\begin{equation}
A(\vec{u})=
\left( \begin{array}{cccc}
   u^3 &  u^4 &  u^1 &  u^2 \\
   u^4 & -u^3 & -u^2 &  u^1 \\
   u^1 &  u^2 & -u^3 & -u^4 \\
   u^2 & -u^1 &  u^4 & -u^3
  \end{array} \right)
\end{equation}
to the path integral calculation of $(\vec{x}_b \mid \vec{x}_a)_E$.
>From the definition of KS transformation (16) and (17), it is easily shown
that
\begin{eqnarray}
x^2 + y^2 + x^2 & = & \left[ (u^1 )^2 + (u^2 )^2 + (u^3 )^2 + (u^4 )^2
\right]^2,  \nonumber \\
\Delta x^2 + \Delta y^2 + \Delta z^2 + \Delta \eta^2 & = & 4 \left[
(u^1 )^2 + (u^2 )^2 + (u^3 )^2 + (u^4 )^2 \right] \nonumber \\
& \times & \left[
(\Delta u^1 )^2 + (\Delta u^2 )^2 + (\Delta u^3 )^2 + (\Delta u^4 )^2
\right],
\end{eqnarray}
$$ \frac{ \partial ( x,y,z,\eta)}{ \partial (u^1 , u^2 , u^3 , u^4 )}  =
2^4 r^2. $$                                                      
Furthermore, the KS transformation of AB potential term is
\begin{eqnarray}
\vec{A} \cdot d\vec{x} & \equiv & \frac{\alpha}{e}
\frac{y dx - x dy}{x^2 + y^2} \nonumber \\
& = & \frac{\alpha}{e}
\left[ \frac{u^1 du^2 - u^2 du^1}{(u^1 )^2 + (u^2 )^2} +
\frac{u^4 du^3 - u^3 du^4}{(u^3 )^2 + (u^4 )^2} \right],
\end{eqnarray}
which makes the system separable like $R^4 \rightarrow R^2 \times R^2$.
This separability from $R^4$ to two independent two-dimensional AB plus
harmonic oscillator systems makes the path-integral calculation of $(\vec{x}_b
\mid \vec{x}_a)_E$
extremely simple.
Although $\vec{u}$-space is noneuclidean with curvature and torsion[12], 
H.Kleinert showed that one can change
\begin{eqnarray}
\int \frac{d \Delta \eta_1}{r_a} 
\int \left[ \prod_{j=2}^{N+1} \left( \frac{M}{2 \pi i \epsilon_s r_j} \right)^2
            d \Delta \vec{x}_j d \Delta \eta_j     \right]  \\ \nonumber 
\Longrightarrow 
\int \frac{d \eta_a}{r_a} 
\int \left[ \prod_{j=1}^{N} \left( \frac{4 M}{2 \pi i \epsilon_s} \right)^2
            d^4 \vec{u}_j  \right]
\end{eqnarray}
without time slicing correction.
After inserting Eq.(20) into (15) and using Eqs.(16, 17, 18), one can
perform the path integral which gives
\begin{eqnarray}
(\vec{x}_b \mid \vec{x}_a)_E&=& 2^{-4} \int_0^{\infty} ds e^{-i \xi s}
\int \frac{d \eta_a}{r_a} 
\left( \frac{4 M \omega}{2 \pi i \sin \omega s} \right)^2 \\ \nonumber
& \times & \sum_{m_1 = -\infty}^{\infty} \sum_{m_2 = -\infty}^{\infty}
e^{i m_1 (\theta_{1,b} - \theta_{1,a})} e^{i m_2 (\theta_{2,b} - \theta_{2,a})}
\\  \nonumber 
& \times & \exp \left[ 2 i M \omega \frac{\cos \omega s}{\sin \omega s} 
( \rho_{1,a}^2 + \rho_{1,b}^2 + \rho_{2,a}^2 + \rho_{2,b}^2 ) \right] \\ 
\nonumber
& \times & I_{|m_1 + \alpha |} \left( \frac{-4 iM\omega}{\sin \omega s}
\rho_{1,a} \rho_{1,b} \right)
I_{|m_2 + \alpha |} \left( \frac{-4 iM\omega}{\sin \omega s}
\rho_{2,a} \rho_{2,b} \right) 
\end{eqnarray}
where $(\rho_1 , \theta_1 )$ and $(\rho_2 , \theta_2 )$ are double polar
coordinates defined by
\begin{eqnarray}
u^1 & = & \rho_1 \sin \theta_1 \nonumber \\
u^2 & = & \rho_1 \cos \theta_1 \nonumber \\
u^3 & = & \rho_2 \cos \theta_2 \\
u^4 & = & \rho_2 \sin \theta_2 \nonumber
\end{eqnarray}
and
\begin{equation}
\omega^2 = - \frac{E}{2M}.
\end{equation}
In order to perform the $\eta_a$-integration we express $(\rho_1 ,\rho_2 ,
\theta_1 , \theta_2 )$ in terms of three-dimensional spherical coordinate
with an auxiliary angle $\gamma$:
\begin{eqnarray}
\rho_1 & = & \sqrt{r} \cos \frac{\theta}{2} \nonumber \\
\theta_1 & = & \frac{ \phi + \gamma + \pi}{2} \nonumber \\
\rho_2 & = & \sqrt{r} \sin \frac{\theta}{2} \\
\theta_2 & = & \frac{\phi - \gamma}{2}. \nonumber
\end{eqnarray}
Then one can change the $\eta_a$-integration into the $\gamma_a$-integration
whose result is easily represented as the Kronecker delta
$\delta_{m_1 ,m_2 }$.
Hence, one can carry out $m_2$-summation and finally
$(\vec{x}_b \mid \vec{x}_a)_E$ becomes 
\begin{eqnarray}
(\vec{x}_b \mid \vec{x}_a)_E &=& \frac{-iM^2 \omega}{\pi}
\sum_{m=-\infty}^{\infty} e^{im (\phi_b - \phi_a )} \nonumber \\
& \times & \int^{\infty}_{0}\frac{dq}{\sinh^2 q} e^{-\frac{\xi}{\omega} q -
2M\omega (r_a + r_b ) \coth q} \\
& \times & I_{|m+\alpha|} \left( \frac{4M \omega}{\sinh q}
\sqrt{r_a r_b} \cos \frac{\theta_a}{2} \cos \frac{\theta_b}{2} \right)
I_{|m+\alpha|} \left( \frac{4M \omega}{\sinh q}
\sqrt{r_a r_b} \sin \frac{\theta_a}{2} \sin \frac{\theta_b}{2} \right).
\nonumber
\end{eqnarray}
In order to obtain the bound-state spectrum one has to check the pole of
$(\vec{x}_b \mid \vec{x}_a)_E$ carfully. Following the appendix of
Ref.[13], one can derive the following integral representation.
\begin{eqnarray}
& & I_{|m+\alpha|} \left( \frac{4M \omega \sqrt{r_a r_b}}{\sinh q}
\cos \frac{\theta_a}{2} \cos \frac{\theta_b}{2} \right)
I_{|m+\alpha|} \left( \frac{4M \omega \sqrt{r_a r_b}}{\sinh q}
\sin \frac{\theta_a}{2} \sin \frac{\theta_b}{2} \right)
\nonumber \\
& = & e^{-i \frac{\pi}{2} |m+ \alpha|} 2^{-3 |m+\alpha|} \pi^{-\frac{1}{2}}
\left[ \Gamma (\frac{1}{2} + |m+ \alpha| ) \right]^{-1}
\left( \frac{4M\omega \sqrt{r_a r_b}}{\sinh q} \sin \theta_a \sin \theta_b
\right)^{|m+ \alpha|}  \nonumber \\
& \times & \int_{0}^{\pi} d\phi (\sin \phi)^{2 |m+ \alpha|}
\left[ \cos \frac{\delta}{2} \right]^{- |m+ \alpha|}
J_{|m+ \alpha|} \left( i \frac{4M\omega \sqrt{r_a r_b}}{\sinh q}
\cos \frac{\delta}{2} \right)
\end{eqnarray}
where $\cos \delta \equiv \cos \theta_a \cos \theta_b
+ \sin \theta_a \sin \theta_b \cos \phi$. If one expands $J_{|m+ \alpha|}$
in Eq.(26) as[13]
\begin{eqnarray}
& & J_{|m+ \alpha|} \left( \frac{4iM \omega \sqrt{r_a r_b}}{\sinh q}
\cos \frac{\delta}{2} \right) \nonumber \\
& = & \sqrt{2} \pi \frac{\Gamma (1+ 2 |m+ \alpha|)}{\Gamma (1+ |m+\alpha|)}
\left( \cos \frac{\delta}{2} \right)^{|m+ \alpha|}
\left( \frac{4iM \omega \sqrt{r_a r_b}}{\sinh q} \right)^{-|m+\alpha|-1} \\
& \times & \sum_{n=0}^{\infty} \sum_{l=0}^{n}
\Bigg[ \frac{ (1+ 2 |m+ \alpha| +2n) (2 |m+ \alpha| +2l) (n-l)!}
{\Gamma (\frac{1}{2}+ |m+ \alpha|) \Gamma(1+ 2 |m+\alpha|+n+l)} \nonumber \\
& \times & J_{1+2 |m+ \alpha|+ 2n}
\left( \frac{4iM\omega \sqrt{r_a r_b}}{\sinh q} \right)
( \sin \theta_a \sin \theta_b )^{-|m+ \alpha|}
(\sin \phi)^{\frac{1}{2} -|m+ \alpha|} \nonumber \\
& \times & P_{n+|m+\alpha|}^{l+ |m+\alpha|} (\cos \theta_a )
  P_{n+|m+\alpha|}^{l+ |m+\alpha|} (\cos \theta_b )
  P_{n+|m+\alpha|- \frac{1}{2}}^{|m+\alpha|- \frac{1}{2}} (\cos \phi ) \Bigg],
\nonumber
\end{eqnarray}
where $P_{n}^{m}$ is a usual associated Legendre function, the
$\phi$-integration
in Eq.(26) is easily performed by using an integral
formula[14]
\begin{eqnarray}
&  & \int_{-1}^{1} dx (1-x^2 )^{\lambda -1} P_{\nu}^{\mu}(x)  \nonumber \\
& = & \frac{ \pi 2^{\mu} \Gamma (\lambda + \frac{\mu}{2})
\Gamma (\lambda - \frac{\mu}{2})}
{ \Gamma (\lambda + \frac{\nu}{2}+ \frac{1}{2})
\Gamma (\lambda - \frac{\nu}{2})
\Gamma (- \frac{\mu}{2} + \frac{\nu}{2} + 1)
\Gamma (- \frac{\mu}{2} - \frac{\nu}{2} + \frac{1}{2} ) }.
\end{eqnarray}
Hence, Eq.(26) is reduced to 
\begin{eqnarray}
& & I_{|m+\alpha|} \left( \frac{4M\omega \sqrt{r_a r_b}}{\sinh q}
\cos \frac{\theta_a}{2} \cos \frac{\theta_b}{2} \right)
I_{|m+\alpha|} \left( \frac{4M\omega \sqrt{r_a r_b}}{\sinh q}
\sin \frac{\theta_a}{2} \sin \frac{\theta_b}{2} \right) \nonumber \\
& = & e^{-i\pi \left( |m+ \alpha|+ \frac{1}{2} \right) }
\frac{\pi \sinh q}{2M \omega \sqrt{r_a r_b}} \\
& \times & \sum_{n=0}^{\infty} \sum_{l=0}^{n}
\Bigg[
      \frac{\sin \frac{l \pi}{2}}{\frac{l \pi}{2}}
\frac{(1+ 2|m+ \alpha| +2n) (|m+ \alpha| + l) (n-l)!}
{ \Gamma(1+ 2|m+\alpha| + n+l) \Gamma(1+ |m+ \alpha| + \frac{l}{2})
\Gamma (1- |m+\alpha| -\frac{l}{2}) } \nonumber \\
& \times & J_{1+ 2|m+ \alpha| + 2n}
\left( \frac{4iM \omega \sqrt {r_a r_b}}{\sinh q} \right)
P_{n+ |m+\alpha|}^{l+|m+\alpha|} (\cos \theta_a )
P_{n+ |m+\alpha|}^{l+|m+\alpha|} (\cos \theta_b ) \Bigg]. \nonumber
\end{eqnarray}
By inserting Eq.(29) into Eq.(25) and after doing
the $q$-integration the final form of the fixed-energy amplitude is 
\begin{eqnarray}
 (\vec{x}_b \mid \vec{x}_a)_E
&=& \frac{1}{8 \omega r_a r_b} \sum_{m=- \infty}^{\infty}
\sum_{n=0}^{\infty} \sum_{l=0}^{n}
\Bigg[
      e^{ im (\phi_b - \phi_a )}
e^{- i \pi (|m+ \alpha| + \frac{1}{2})} \frac{2}{l \pi}
\sin \frac{l\pi}{2} \nonumber \\
& \times & \frac{ (|m+ \alpha| +l) (n-l)!
\Gamma(1+ |m+\alpha| +n +\frac{\xi}{2\omega}) }
{ \Gamma(1+ 2|m+\alpha|+n+l) \Gamma(1+ 2|m+\alpha|+ 2n)
  \Gamma(1- |m+\alpha|- \frac{l}{2}) \Gamma(1+ |m+\alpha|+ \frac{l}{2}) }
\nonumber \\
& \times & P_{n+ |m+\alpha|}^{l+ |m+\alpha|} (\cos \theta_a )
P_{n+ |m+\alpha|}^{l+ |m+\alpha|} (\cos \theta_b ) \\
& \times & W_{ -\frac{\xi}{2\omega}, \frac{1+ 2|m+\alpha|+2n}{2} }
\left( 4M\omega \mbox{Max} (r_a ,r_b ) \right)
M_{ \frac{\xi}{2\omega}, \frac{1+ 2|m+\alpha|+2n}{2} }
\left(- 4M\omega \mbox{Min} (r_a ,r_b ) \right) \Bigg], \nonumber
\end{eqnarray}
where $W_{a,b}$ and $M_{a,b}$ are usual Whittaker functions.

The energy spectrum of ABC system is deduced from the poles of the
Gamma function in numerator:
\begin{equation}
E_{n,n',m} = - \frac{ M \xi^2} {2( 1+ |m+\alpha| + n+n' )^2},
\hspace{0.5cm}(n,n'=0,1,2, \cdots).
\end{equation}
Our result on the energy spectrum agrees with those of Ref.[5]
and Ref.[6].

\indent In summary, we derived the exact propagator of ABC system by using
pseudotime method and dimensional extension technique firstly used by
Duru and Kleinert in Ref.[8]. Also the bound state energy spectrum
is deduced. It might be straightforward to derive the exact propagator of
two-dimensional ABC system if one use a Levi-Civita transformation which is a
two-dimensional version of the KS transformation. If the propagator of two
dimensional ABC system is derived, one can derive the propagator of 
spin-1/2 ABC system by incorporating the self-adjoint extension method
into the path integral formalism which was suggested in Ref.[15].
The evaluation of spin-1/2 ABC propagator might be very helpful for the
analysis of time-dependent anyon scattering and study on the statistical
properties of anyon system. This work is in progress.

\end{document}